\documentclass[prb,showpacs,twocolumn,aps,superscriptaddress,a4paper]{revtex4-1}
\usepackage{dcolumn}
\usepackage{amsmath}
\usepackage{graphicx}
\usepackage{latexsym}
\usepackage{amsfonts}
\usepackage{amssymb}
\usepackage{color}

\newcommand{\be}{\begin{equation}}
\newcommand{\beq}{\begin{equation}}
\newcommand{\ee}{\end{equation}}
\newcommand{\bea}{\begin{eqnarray}}
\newcommand{\eea}{\end{eqnarray}}
\newcommand{\vn} {{\bf n}}
\newcommand{\vv} {{\bf v}}
\newcommand{\nn} {\nonumber}
\newcommand{\tr}[1] { {\rm Tr} \left\{ #1 \right\} }

\begin{document}

\title{Lattice density functional theory at finite temperature with
strongly density-dependent exchange-correlation potentials}

\author{Gao Xianlong}
\affiliation{Department of Physics, Zhejiang Normal University, Jinhua 321004,
China}
\author{A-Hai Chen}
\affiliation{Department of Physics, Zhejiang Normal University, Jinhua 321004,
China}
\author{I. V. Tokatly}
\affiliation{Nano-Bio Spectroscopy Group and European Theoretical Spectroscopy
Facility (ETSF) Scientific Development Centre,
Dpto. de F\'{i}sica de Materiales,
Universidad del Pa\'{i}s Vasco UPV/EHU,
Av. Tolosa 72, E-20018 San Sebasti\'{a}n, Spain}
\affiliation{IKERBASQUE, Basque Foundation for Science, E-48011 Bilbao, Spain}

\author{S. Kurth}
\affiliation{Nano-Bio Spectroscopy Group and European Theoretical Spectroscopy
Facility (ETSF) Scientific Development Centre,
Dpto. de F\'{i}sica de Materiales,
Universidad del Pa\'{i}s Vasco UPV/EHU,
Av. Tolosa 72, E-20018 San Sebasti\'{a}n, Spain}
\affiliation{IKERBASQUE, Basque Foundation for Science, E-48011 Bilbao, Spain}

\date{\today}

\begin{abstract}
The derivative discontinuity of the exchange-correlation (xc) energy at integer
particle number is a property of the exact, unknown xc functional of density
functional theory (DFT) which is absent in many popular local and semilocal
approximations. In lattice DFT, approximations exist which exhibit a
discontinuity in the xc potential at half filling. However, due to convergence
problems of the Kohn-Sham (KS) self-consistency cycle, the use of these
functionals is mostly restricted to situations where the local density is away
from half filling. Here a numerical scheme for the self-consistent solution
of the lattice KS Hamiltonian with a local xc potential with rapid (or
quasi-discontinuous) density dependence is suggested. The problem is
formulated in terms of finite-temperature DFT where the discontinuity in the
xc potential emerges naturally in the limit of zero temperature. A simple
parametrization is suggested for the xc potential of the uniform 1D
Hubbard model at finite temperature which is obtained from the solution of the
thermodynamic Bethe ansatz. The feasibility of the numerical scheme is
demonstrated by application to a model of fermionic atoms in a harmonic trap.
The corresponding density profile exhibits a plateau of integer occupation at
low temperatures which melts away for higher temperatures.
\end{abstract}


\maketitle

\section{Introduction}

Originally, static (ground-state) density functional theory (DFT) has been
formulated \cite{HohenbergKohn:64,KohnSham:65} for many-electron systems in
the continuous space of three spatial dimensions with the electrons
interacting via the Coulomb interaction. On the other hand, many phenomena in
many-particle physics are studied in terms of model systems on discrete
lattices, typically of tight-binding form. The one-dimensional Hubbard model
or the Anderson impurity model are just the most prominent examples.
Typically, these models are studied with techniques different from DFT.
However, DFT can be a useful tool for the investigation of these models as
well, especially when one wants to take into account the effects of
non-uniform external potentials\cite{GunnarssonSchoenhammer:86,LimaSilvaOliveiraCapelle:03,XianlongPoliniTosiCampoCapelleRigol:06,MaPilatiTroyerDai:12}.
For example, cold atoms in optical lattices confined in an harmonic trap
may very well be modelled by a lattice model with confining external potential
where the particles interact through a Hubbard-type interaction~\cite{Review}.

The idea of formulating DFT for electrons on a lattice of sites has been
pioneered by Sch\"onhammer and coworkers
\cite{GunnarssonSchoenhammer:86,SchoenhammerGunnarssonNoack:95}. As with
usual DFT, the applicability and success of lattice DFT hinges on the
availability of approximations to the unknown exchange-correlation (xc)
functional. Capelle and coworkers proposed a local functional for the
xc energy per site based on the Bethe-ansatz solution of the uniform 1D
Hubbard model at zero temperature
\cite{LimaSilvaOliveiraCapelle:03,SilvaLimaMalvezziCapelle:05}. Thus, for
one-dimensional lattice models, the 1D Hubbard model takes the role
of the uniform electron gas in continuum formulation of DFT as an exactly
solvable model system which provides the essential input for the
construction of the local approximation. In the same spirit, xc functionals
have been suggested based on other Hubbard lattice models such as the
two-dimensional hexagonal lattice \cite{IjaesHarju:10} or the simple cubic
lattice in 3D \cite{KarlssonPriviteraVerdozzi:11}.

An interesting property of the Bethe-ansatz local density approximation (BALDA) in 1D
(and also its counterparts for 2D or 3D lattice models) is its
discontinuous form of the xc potential at half filling or integer occupation
\cite{LimaOliveiraCapelle:02}. Physically, this discontinuity is a direct
consequence of the Mott-Hubbard gap of the Hubbard model while in the DFT
context it is nothing but the well-known derivative discontinuity of the xc
energy at integer particle number at zero temperature
\cite{PerdewParrLevyBalduz:82}.

The BALDA has been successfully applied to spatially inhomogeneous
Hubbard supperlattices \cite{SilvaLimaMalvezziCapelle:05}, to
cold fermionic atoms in a harmonic trap, both with repulsive
\cite{XianlongPoliniTosiCampoCapelleRigol:06} and attractive
\cite{XianlongRizziPoliniFazioTosiCampoCapelle:07} electronic interaction
and to the study of the static and dynamic linear density response
\cite{SchenkDzierzawaSchwabEckern:08,AkandeSanvito:10}. Extensions of the
BALDA have been suggested to systems in static magnetic fields
\cite{AkandeSanvito:12} and, in the adiabatic form,
to the domain of time-dependent DFT \cite{Verdozzi:08} where it has been
used to study the dynamics of finite Hubbard clusters. Recently the adiabatic
BALDA has been applied to describe the time evolution of trapped 1D lattice
fermions in the Mott insulator regime
\cite{KarlssonVerdozziOdashimaCapelle:11}. A modified version
of the BALDA has been used in the study of time-dependent transport through
an Anderson impurity \cite{KurthStefanucciKhosraviVerdozziGross:10} where the
discontinuity has been related to Coulomb blockade.

From a physical point of view the discontinuity at integer particle number
is certainly a desirable
property for an approximation to have, at least at zero temperature. As we
will argue below, the zero-temperature discontinuity may be viewed as the
zero temperature limit of a {\em continuous} xc potential at finite
temperature. From a practical point of view, if a local discontinuous (or
rapidly varying) xc potential is used, one often faces convergence problems
of the Kohn-Sham (KS) self-consistency cycle
\cite{XianlongPoliniTosiCampoCapelleRigol:06} essentially whenever the
local density is close to integer occupation.

In the present work we propose a practical solution to this convergence
problem by viewing it as an equivalent problem of finding the solution to a
coupled set of nonlinear equations. In Sec.~\ref{KS_problem}, we
start with a general discussion of the KS self-consistency cycle and possible
convergence problems when using KS potentials which vary rapidly for small
variations in the density. The problem is illustrated explicitly on the
simple, exactly soluble model system of a single interacting site in contact
with a heat and particle bath. In Sec.~\ref{lattice_DFT}, we then introduce
the 1D lattice models studied throughout this work and briefly summarize
the idea of a local approximation for lattice models which has been discussed
in the literature. We will
work in the framework of finite-temperature DFT \cite{Mermin:65},
and Sec.~\ref{lda_fintemp} is devoted to the construction of an
approximate xc potential for this framework. We construct the
xc potential of the uniform 1D Hubbard model for {\em finite} temperatures
based on the thermodynamic Bethe ansatz. We provide a
simple parametrization of this potential using insights gained from the simple
single site model discussed earlier.
In Sec.~\ref{conv} we introduce our algorithm for the practical
solution of the self-consistency problem which is based on a multi-dimensional
bisection method.
In Sec.~\ref{applic} we show a numerical application of the method to the
problem of interacting particles in a harmonic trap before we present our
conclusions in Sec.~\ref{conclus}. In the Appendix
we provide explicit expressions for the xc free energy per site for the
simple parametrization of the thermodynamic Bethe ansatz solutions to the
uniform Hubbard model.

\section{Kohn-Sham problem with rapidly varying density functionals}
\label{KS_problem}

The implementation of DFT via the KS method gained enormous popularity
because it reduces calculations of the density $n({\bf r})$ in a complicated
strongly interacting system to computing $n({\bf r})$ for a reference system of
noninteracting KS particles. The KS particles move in the presence of an
effective potential $v^{\rm KS}= v + v_{\rm Hxc}[n]$, where $v$ is an external
potential and $v_{\rm Hxc}[n]$ is the Hartree-exchange-correlation (Hxc)
potential which
depends on the density and is adjusted self-consistently to reproduce the
physical density distribution of the interacting system. The self-consistent
nature of the KS problem makes it nonlinear and thus not absolutely trivial.
In fact, the whole point of the present paper is to identify one of the
potentially dangerous physical situations and to propose a recipe for its
solution.

\subsection{KS self-consistency as a fixed point problem: the issue of
convergence}
\label{fixed_point}

Assuming that the potential $v_{\rm Hxc}[n]$ as functional of the density is
known, the general KS problem can be formulated as follows. We have to find a
set of KS orbitals $\varphi^{(\alpha)}$ and KS energies $\varepsilon_{\alpha}$
by solving a one-particle stationary Schr\"odinger equation
\begin{equation}
 \label{ks_eq_general}
 (\hat{t} + v + v_{\rm Hxc}[n])\varphi^{(\alpha)} =
\varepsilon_{\alpha}\varphi^{(\alpha)},
\end{equation}
where $\hat{t}$ is the one-particle kinetic energy operator. As the operator
in Eq.~(\ref{ks_eq_general}) depends on the density we need an additional
``self-consistency equation'' that relates the set of
$\{\varphi^{(\alpha)},\varepsilon_{\alpha}\}$ to $n({\bf r})$. Obviously this
equation is simply the standard definition of the density of noninteracting
particles
\begin{equation}
 \label{density_def}
n({\bf r}) = 2 \sum_{\alpha} f(\varepsilon_{\alpha})\left|\varphi^{(\alpha)}({\bf r})\right|^2,
\end{equation}
where the factor two comes from spin.
$f(\omega) = (1 + \exp(\beta (\omega-\mu)))^{-1}$ is the Fermi distribution,
$\beta=1/T$ is the inverse temperature, and $\mu$ is the chemical potential
which is either given externally or determined by fixing the total number of
particles.

Calculation of the density from Eqs.~(\ref{ks_eq_general})-(\ref{density_def})
is equivalent to finding a fixed point of a certain density functional.
Indeed, the eigenvalue problem of Eq.~(\ref{ks_eq_general}) defines a map
$n\mapsto\{\varphi^{(\alpha)},\varepsilon_{\alpha}\}$ from the density to the
set of KS eigenfunctions and eigenvalues, i.~e., it determines the functionals
$\varphi^{(\alpha)}[n]$ and $\varepsilon_{\alpha}[n]$. Inserting these
functionals into Eq.~(\ref{density_def}) we obtain the following form
\begin{equation}
 \label{fixed_point_eq}
 n = 2 \sum_{\alpha} f(\varepsilon_{\alpha}[n])\left|\varphi^{(\alpha)}[n]\right|^2 \equiv G[n],
\end{equation}
which is a typical fixed point problem for the functional $G[n]$ on the right
hand side.

In practice, the self-consistent KS problem of
Eqs.~(\ref{ks_eq_general})-(\ref{density_def}), or equivalently the fixed
point problem of Eq.~(\ref{fixed_point_eq}), is commonly solved iteratively.
In the simplest case one starts with some initial guess $n^{(0)}$ for the
density and constructs a sequence of iterations $n^{(k)}$ as follows
\begin{equation}
 \label{ks_iterations}
 n^{(0)},\, n^{(1)} =G[n^{(0)}],\dots,\, n^{(k)} =G[n^{(k-1)}],\dots
\end{equation}
The limiting point of this sequence presumably gives a self-consistent 
solution of the KS equation
\begin{equation}
 \label{limit}
 n = \lim_{k\to\infty} n^{(k)}.
\end{equation}
Unfortunately the assumed convergence cannot be guaranteed in general,
in spite of the fact that the original KS problem definitely has a unique
solution. From the Banach fixed point theorem (the contraction mapping
principle) we know that the sequence of Eq.~(\ref{ks_iterations}) does
necessarily converge to a unique fixed point if the functional $F[n]$ is
contractive, i.~e., if the following condition is satisfied
\begin{equation}
 \label{contraction}
 \| G[n]-G[n']\| \le \lambda \| n - n'\|, \quad 0<\lambda < 1,
\end{equation}
where $\|\dots\|$ means a properly chosen norm in the space of densities.
Apparently this condition requires $F[n]$ to be a sufficiently smooth
functional of the density, which is not always the case. Moreover there are
important physical situations where the inequality of Eq.~(\ref{contraction})
is always violated. To understand this more clearly we estimate the left hand
side of Eq.~(\ref{contraction}) for a small density variation $n'=n+\delta n$
with $\delta n \ll n$
\begin{equation}
 \label{delta_F}
 \| G[n]-G[n']\| \approx \| \chi\frac{\delta v_{\rm Hxc}}{\delta n} (n - n')\|
\end{equation}
where $\chi$ is the density response function. Obviously the right hand side
of Eq.~(\ref{delta_F}) cannot be smaller than $\lambda \| n - n'\|$ with
$\quad 0<\lambda < 1$ if the Hxc potential is a rapidly varying functional of
$n$, i.~e., if $\frac{\delta v_{\rm Hxc}}{\delta n}$ is large at least for some
directions in density space. Physically this should always happen in
systems composed of weakly coupled fragments if the number of
particles in at least one of the fragments is close to an integer value. Then
a density transfer to/from this fragment causes a strong variation of the
potential. The origin of this behavior is in the famous discontinuity of the
exact xc potential at integer number of
particles~\cite{PerdewParrLevyBalduz:82}. The most prominent examples
of systems demonstrating such a behavior are molecules close to dissociation or
strongly correlated solids near the Mott-Hubbard transition. In all those
systems where the physics is governed by a nearly discontinuous xc potential
the standard iterative procedure of solving the KS equations will not converge.

In the next subsection we explicitly illustrate the above general argument
by considering a very simple model system -- a single lattice site which can
host at most two spin-1/2 fermions. The purpose for studying this model is
twofold. Firstly, this is probably the only case where the exact xc potential
can be found analytically for any temperature. The corresponding KS problem
possesses an analytic solution and, because of its simple structure, clearly
shows when and why the existing unique fixed point cannot be reached
iteratively. Secondly, a single site DFT serves as a paradigmatic example for
more general interacting lattice models. In fact, the analytic form of the
single site Hxc potential will later be used to construct a simple
parametrization for the Hxc potential of the uniform Hubbard model at finite
temperatures.

\subsection{KS-DFT for a single site model}
\label{SSM}

Let us consider one single-orbital site in contact with a heat and particle
bath at inverse temperature $\beta$ and chemical potential $\mu$
\cite{StefanucciKurth:11,EversSchmitteckert:11}.

The Hamiltonian for this single site model (SSM) in the presence of an on-site
interaction is given by
\be
\hat{H}_{\rm SSM} = v_0 \hat{n}_0 + U \hat{n}_{0,\uparrow} \hat{n}_{0,\downarrow}
\ee
where $v_0$ is the on-site energy and $U$ is the charging energy, 
$\hat{n}_{0,\sigma}$ and
$\hat{n}_{0} = \sum_{\sigma=\uparrow,\downarrow} \hat{n}_{0,\sigma}$
are the operators for the on-site density with spin $\sigma$ and for the
total density, respectively. Similarly, for the non-interacting case the 
single-site Hamiltonian reads
\be
\hat{H}_{\rm SSM}^s = v_s \hat{n}_0
\ee
with on-site energy $v_s$. The complete Fock space of both Hamiltonians is
spanned by the states $|0\rangle$, $|\uparrow \rangle$, $|\downarrow \rangle$,
and $|\uparrow \downarrow \rangle$ with particle
occupation of zero, one, and two. These states are both eigenstates of
$\hat{H}_{\rm SSM}$ with eigenvalues $0$, $v_0$, $v_0$, and $2v_0+U$,
as well as eigenstates of $\hat{H}_{\rm SSM}^s$ with eigenvalues $0$, $v_s$,
$v_s$, and $2v_s$, respectively. For the single site model, the particle number
operator is equal to the density operator, $\hat{N}=\hat{n}_0$, and the density
$n_0 = \tr{\hat{\rho} \hat{n}_0}$ for the interacting case then reads
\bea
\lefteqn{
n_0 = }\nn\\
&& \frac{ 2 \exp(-\beta (v_0-\mu)) + 2
\exp(-\beta (2(v_0-\mu)+U)) }{Z^{\rm SSM}}
\label{ssm_dens_int}
\eea
where
\be
Z^{\rm SSM} = 1 + 2 \exp(-\beta (v_0-\mu)) + \exp(-\beta (2(v_0-\mu)+U))
\label{ssm_partfunc_int}
\ee
is the grand canonical partition function. Eq.~(\ref{ssm_dens_int}) only
depends on the
quantity $\tilde{v}_0 = v_0-\mu$ and the function $n_0(\tilde{v}_0)$ can be
inverted explicitly leading to
\be
\tilde{v}_0(n_0) = - U - \frac{1}{\beta} \ln\left( \frac{\delta n +
\sqrt{\delta n^2 + e^{-\beta U} ( 1 - \delta n^2)}}{1 - \delta n}\right)
\label{ssm_pot_int}
\ee
where $\delta n = n_0 -1$.

Following the same lines, for the non-interacting case the density reads
\be
n_0^s = \frac{2 \exp(-\beta (v_s-\mu)) + 2
\exp(-\beta (2(v_s-\mu)))}{Z_s^{\rm SSM}}
\label{ssm_dens_nonint}
\ee
with the non-interacting partition function
\be
Z_s^{\rm SSM} = 1 + 2 \exp(-\beta (v_s-\mu)) + \exp(-\beta (2(v_s-\mu))) \; .
\label{ssm_partfunc_nonint}
\ee
Again, the density $n_0^s$ only depends on the quantity $\tilde{v}_s= v_s -
\mu$ and one can invert $n_0^s(\tilde{v}_s)$ to yield
\be
\tilde{v}_s(n_0^s) = - \frac{1}{\beta} \ln\left( \frac{1+ \delta n^s}
{1 - \delta n^s}\right)
\label{ssm_pot_nonint}
\ee
with $\delta n^s = n_0^s - 1$.

The exact Hxc potential for the SSM can now easily be calculated by requiring
that the interacting density equals the non-interacting one $n_0=n_0^s=:n$
and taking the difference of the two expressions (\ref{ssm_pot_nonint}) and
(\ref{ssm_pot_int}), i.e.,
\be
v_{\rm Hxc}^{\rm SSM}(n,U,T) = \tilde{v}_s(n) - \tilde{v}_0(n) = \frac{U}{2} +
g(n-1)
\label{v_hxc_ssm}
\ee
where
\be
g(x) = \frac{U}{2} + \frac{1}{\beta} \ln \left( \frac{x +
\sqrt{x^2 + e^{-\beta U} ( 1 - x^2)}}{1 + x}\right)
\label{gfunc_ssm}
\ee
which is easily shown to be an odd function of its argument, $g(-x)=-g(x)$.

\begin{figure}[t]
\includegraphics[width=0.47\textwidth]{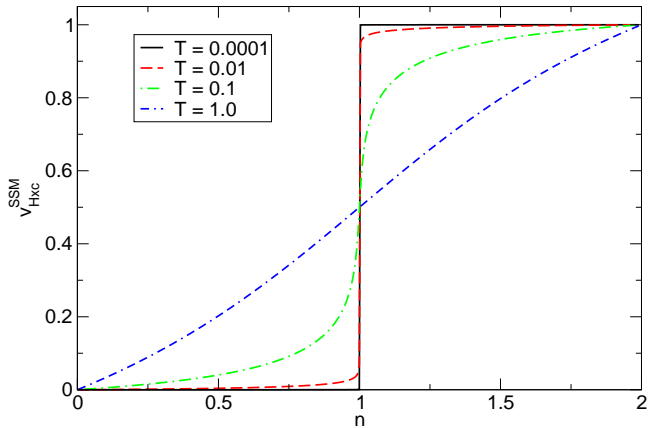}
\caption{Hartree-exchange-correlation potential of the single-site model for
different temperatures $T=1/\beta$. Energies given in units of $U$.}
\label{vhxc_ssm}
\end{figure}

In Fig.~\ref{vhxc_ssm} we show the SSM Hxc potential $v_{\rm Hxc}^{\rm SSM}(n)$
as function of the density for different temperatures. At low temperatures,
$v_{\rm Hxc}^{\rm SSM}(n)$ becomes an extremely rapidly varying function of
$n$ in the vicinity of $n=1$, approaching a step function with a step of
height $U$ at $n=1$ in the limit of zero temperature.

Now, having at hand the exact Hxc potential, we can study the KS problem. In
the single site DFT, the general fixed point equation (\ref{fixed_point_eq})
reduces to the following algebraic transcendental equation
\begin{equation}
 \label{fixed_point_ssm}
 n = 2 f\left(v_0 + v_{\rm Hxc}^{\rm SSM}(n)\right) \equiv G(n).
\end{equation}
\begin{figure}[t]
\includegraphics[width=0.47\textwidth]{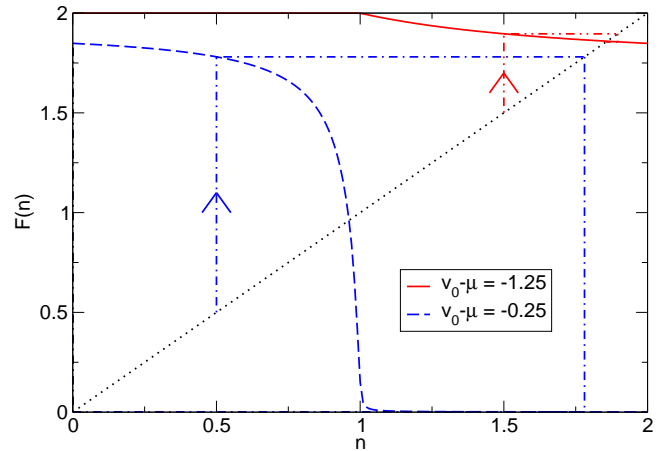}
\caption{Left-hand side (dotted black line ) and r.h.s. of
Eq.~(\ref{fixed_point_ssm}) for
different values of $v_0-\mu$ and temperature $T=0.1$. (Energies given in
units of $U$.) For given $v_0-\mu$, the self-consistent solution is given by
the intersection of the corresponding $G(n)$ with the straight line. For
$-U<v_0-\mu<0$, the iterative scheme $n^{(k+1)}=G(n^{(k)})$, indicated by the
dash-dotted lines, does not converge.
}
\label{ssm}
\end{figure}
This equation can be easily solved analytically. By construction, the solution
to Eq.~(\ref{fixed_point_ssm}) simply returns the function $n(T,\mu)$ defined
by Eqs.~(\ref{ssm_dens_int}) and (\ref{ssm_partfunc_int}). In
Fig.~\ref{ssm} we show the left (straight line) and right hand sides of
Eq.~(\ref{fixed_point_ssm}) for two different values of $v_0-\mu$. Obviously,
for a given $v_0$ there is only one intersection between $n$ and $G(n)$, which
means that the function $G(n)$ always has only one fixed point. However, if the
expected solution lies in the region of fast variation of $v_{\rm Hxc}$, i.e.
$n\sim 1$, the fixed point cannot be reached by iterations. Independently of
the choice of the initial guess, after a few iterations we enter a limiting
cycle with the density endlessly jumping between $n\approx 0$ and
$n\approx 2$. This behavior is generic for low enough temperatures, $T\ll U$,
and the chemical potential in the region $v_0<\mu<v_0+U$, which are the
conditions ensuring that $v_{\rm Hxc}(n)$ has a step-like form (see
Fig.~\ref{vhxc_ssm}) and the physical on-site occupation is close to unity.
Examples of the iterative cycle are indicated in Fig.~\ref{ssm} showing
the convergence of the cycle of Eq.~(\ref{ks_iterations}) for
$v_0-\mu = -1.25$, and the lack of convergence for $v_0-\mu = -0.25$.

From the discussion in Sec.~\ref{fixed_point} it is clear that the same type
of non-convergence of the KS iterative sequence should occur in any system
with a discontinuous/rapidly varying xc potential. In the rest of this paper
we study and solve this problem for lattice models where the discontinuity
of $v_{\rm Hxc}$ reflects Mott-Hubbard correlations and can easily be captured
at the level of a local density approximation.

\section{Lattice density functional theory}
\label{lattice_DFT}

\subsection{Lattice DFT: Formalism and Model}
\label{L-DFT_general}

As a particular example of a lattice model, we consider one-dimensional,
interacting many-electron
systems on a tight-binding lattice described by the Hamiltonian
\bea
\hat{H} &=& - t \sum_{i=1}^{N_s-1} \sum_{\sigma=\uparrow,\downarrow} \left(
\hat{c}_{i,\sigma}^{\dagger} \hat{c}_{i+1,\sigma} + H.c. \right) \nn\\
&&+ U \sum_{i=1}^{N_s} \hat{n}_{i,\uparrow} \hat{n}_{i,\downarrow} + \sum_{i=1}^{N_s} v_i \hat{n}_{i}
\label{hamil_int}
\eea
where $\hat{c}_{i,\sigma}^{\dagger}$ ($\hat{c}_{i,\sigma}$) are the fermionic
creation (annihilation) operators for an electron with spin $\sigma$ at site
$i$, $\hat{n}_{i,\sigma}= c_{i,\sigma}^{\dagger} c_{i,\sigma}$ and
$\hat{n}_{i} = \sum_{\sigma=\uparrow,\downarrow} \hat{n}_{i,\sigma}$
are the operators for the density of electrons with spin $\sigma$ and for the
total electron density at site $i$, respectively. The nearest neighbor hopping
element is $t$, $U$ is the Hubbard interaction. $v_i$ is the external
potential at site $i$ and $N_s$ is the total number of sites. For simplicity,
we consider systems in the absence of magnetic fields. For the grand-canonical
ensemble, when the system is in contact with a heat bath at inverse temperature
$\beta$ and a particle bath at chemical potential $\mu$, the statistical
operator is
\be
\hat{\rho} = \frac{1}{Z} \exp(-\beta (\hat{H} - \mu \hat{N}))
\ee
where $\hat{N}= \sum_{i=1}^{N_s} \hat{n}_i$ is the operator for the total
number of particles and the grand canonical partition function is
$Z = \tr{\exp(-\beta (\hat{H} - \mu \hat{N}))}$ with the trace over all states
of Fock space. In thermal equilibrium, an observable described by the
operator $\hat{A}$ then takes the value $A = \tr{ \hat{\rho} \hat{A}}$.

In the spirit of DFT at finite temperatures \cite{Mermin:65}, the Hamiltonian
(\ref{hamil_int}) is mapped onto the following Hamiltonian of non-interacting
electrons
\be
\hat{H}^{\rm KS} = - t \sum_{i=1}^{N_s} \sum_{\sigma=\uparrow,\downarrow} \left(
\hat{c}_{i,\sigma}^{\dagger} \hat{c}_{i+1,\sigma} + H.c. \right)
+ \sum_{i=1}^{N_s} v_i^{\rm KS} \hat{n}_{i}
\label{hamil_ks}
\ee
where the effective single particle KS potential $v^{\rm KS}_i$
at site $i$ is chosen such that the equilibrium density $n_i = \tr{\hat{\rho}
\hat{n}_i}$ of the interacting Hamiltonian (\ref{hamil_int}) and the KS
Hamiltonian (\ref{hamil_ks}) are the same for all sites. The KS potential at
site $i$ then has the form
\be
v_i^{\rm KS} = v_i + v_i^{\rm Hxc}
\label{ks_pot}
\ee
where $v_i$ is the external potential at site $i$ and, similarly,
$v_i^{\rm Hxc}$ is the Hxc potential at site $i$.
In general, the Hxc potential at site $i$ depends on the equilibrium density
at all other sites, i.e., $v_i^{\rm Hxc}=v_i^{\rm Hxc}(\{n_j\})$. Typically,
however, the exact form of the Hxc potential is unknown and one
has to resort to approximations. Once an approximation to $v_i^{\rm Hxc}$ has
been specified, the equilibrium density of the KS Hamiltonian
$\hat{H}^{\rm KS}$ can be found by self-consistent solution of the KS equation
on the lattice
\be
\sum_{j=1}^{N_s}\left( - t_{ij} + v_i^{\rm KS} \delta_{ij} \right)
\varphi_j^{(\alpha)} = \varepsilon_{\alpha} \varphi_i^{(\alpha)}
\label{ks_eq}
\ee
(with $t_{ij} = t$ for $j=i\pm1$ and $t_{ij} = 0$ otherwise)
together with
\be
n_i = 2 \sum_{\alpha} f(\varepsilon_{\alpha}) |\varphi_i^{(\alpha)}|^2 \;.
\label{dens}
\ee

\subsection{Local density approximations in the lattice DFT}
\label{L-DFT_LDA}

In the {\em local} approximation, the Hxc potential at site $i$ only
depends on the density at the same site $i$,
$v_i^{\rm Hxc,loc}=v_{\rm Hxc}^{\rm mod}(n_i)$.
The functional dependence of $v_{\rm Hxc}^{\rm mod}(n)$ on the density is
extracted from some interacting model system for which the exact solution can
be constructed by analytical and/or numerical techniques. Probably the most
prominent example of such a functional for lattice-DFT is the local density
approximation (LDA) based on Bethe-ansatz solution of the uniform Hubbard
model in 1D (Bethe-ansatz LDA, BALDA) at zero temperature
\cite{LimaSilvaOliveiraCapelle:03,SilvaLimaMalvezziCapelle:05,LiebWu:68}.

Strictly speaking, at zero temperature and exactly at half-filling ($n=1$),
the BALDA xc potential is not defined since the xc energy per particle is
not differentiable at this point. One pragmatic way around this mathematical
problem is to smoothen the discontinuity in some ad-hoc manner
\cite{KurthStefanucciKhosraviVerdozziGross:10,KarlssonVerdozziOdashimaCapelle:11}.
Alternatively one can construct
xc functionals for finite temperature which approach a discontinuous function
in the zero temperature limit. We have already followed this route in
Sec.~\ref{SSM} to formulate a single site DFT, and will pursue it
further in Sec.~\ref{lda_fintemp} for the 1D Hubbard model.

Although one can avoid the use of truly discontinuous xc potentials in this
way, the resulting KS potentials will still be very rapidly varying functions
of the density. This is exactly the property leading to a non-convergence of a
simple iterative procedure, a fact which has been recognized in attempts to
use the BALDA xc potential within the usual KS self-consistency cycle
\cite{XianlongPoliniTosiCampoCapelleRigol:06}.

\section{Local approximations at finite temperature}
\label{lda_fintemp}
In the present Section we propose several versions of a local functional at
finite temperature for which the corresponding Hxc potentials exhibit rapid
variations as function of the density. This functional is based on the
thermodynamic Bethe ansatz (TBA) solution of the uniform Hubbard model in one
dimension\cite{Takahashi:72} and is thus an extension of the
corresponding work at zero temperature
\cite{SchoenhammerGunnarssonNoack:95,LimaSilvaOliveiraCapelle:03}.

\begin{figure}[t]
\includegraphics[width=0.47\textwidth]{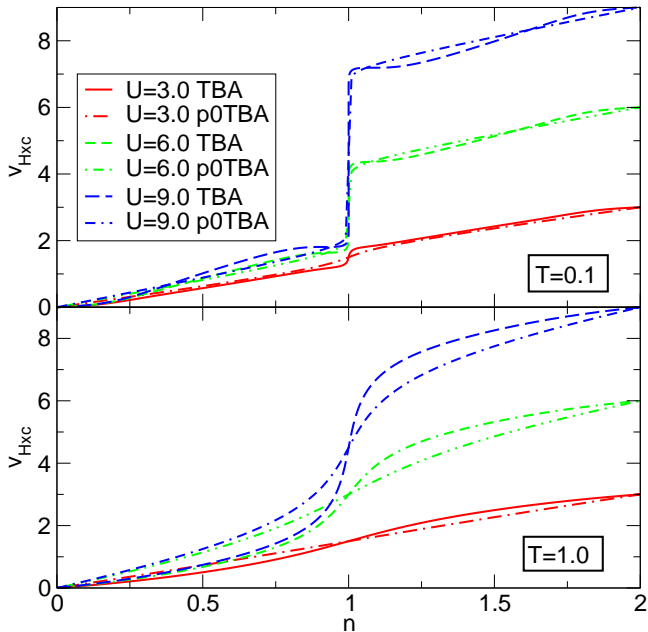}
\caption{Comparison of the fully numerical TBA Hxc potential with p0TBA
parametrization of Eq.~(\protect\ref{vhxc_tba_para0}) for different
interactions and two typical temperatures.}
\label{vhxc_tba_comp}
\end{figure}

We have numerically solved the coupled integral equations of the TBA following
Refs.~\onlinecite{KawakamiUsukiOkiji:89,UsukiKawakamiOkiji:90,TakahashiShiroishi:02}.
For given inverse temperature $\beta$, the density is calculated as a function
of the chemical potential which can be inverted to give the chemical
potential as function of the density.  For the interacting and non-interacting
cases these inverse functions are denoted as $\mu(n)$ and $\mu_s(n)$,
respectively. From these two functions we obtain the density-dependent Hxc
potential of the TBA as
\be
v_{\rm Hxc}^{\rm TBA}(n) = \mu(n) - \mu_s(n)
\label{vhxcTBA}
\ee
In Fig.~\ref{vhxc_tba_comp} we show the density dependence of the TBA Hxc
potentials for various values of the interaction $U$ and various temperatures
$T=1/\beta$. At low temperatures and for sufficiently large values of $U$,
the TBA Hxc potential $v_{\rm Hxc}^{\rm TBA}(n)$ exhibits rapid variations
around half filling ($n=1$) as function of density. In the zero-temperature
limit this feature reduces to a step whose height is given by the Mott-Hubbard
gap. This gap can be expressed in terms of the parameters of the model as
(from now on all energies are given in units of the hopping matrix element $t$
unless otherwise noted)
\be
\Delta_{0}(U) = \frac{16}{U} \int_1^{\infty} {\rm d} x \;
\frac{\sqrt{x^2-1}}{\sinh(2 \pi x /U)}
\label{hubbard_gap}
\ee
which is nothing but the derivative discontinuity of the uniform Hubbard
model at half filling \cite{LimaOliveiraCapelle:02,LiebWu:68}.

Away from $n=1$, even at low temperatures the dependence of
$v_{\rm Hxc}^{\rm TBA}(n)$ on the density is rather slow and smooth. For high
temperatures, the sharp feature around half filling is washed out. As a
consequence of particle-hole symmetry, the TBA Hxc potential takes the value
$U/2$ at $n=1$ for all temperatures and exhibits a point symmetry around this
point as function of density, i.e., $v_{\rm Hxc}^{\rm TBA}(n)=U/2 +
g^{\rm TBA}(n-1)$ with $g^{\rm TBA}(-x)=-g^{\rm TBA}(x)$. Furthermore the values
of the TBA Hxc potential at the endpoints of the density interval are
$v_{\rm Hxc}^{\rm TBA}(0)=0$ and $v_{\rm Hxc}^{\rm TBA}(2)=U$ for all temperatures.

We use these observations to design a hierarchy of analytic parametrizations
of the fully numerical TBA Hxc potential which can easily be used in practical
calculations. In the construction of these parametrization we make use of the
simple analytic form of the Hxc potential of the single site model discussed
in Sec.~\ref{SSM}.

\subsection{Lowest level single-site-motivated parametrization of the numerical TBA: p0TBA}

Our simplest functional is aimed at reproducing the main qualitative features
of the full numerical TBA based Hxc potential. These are the point symmetry of
the function  $v_{\rm Hxc}^{\rm TBA}(n)$ (reflecting the electron-hole
symmetry), and the step structure at $n=1$, which gradually washes out at
higher temperatures.

Precisely this pattern is also observed in the Hxc potential of the single-site
model: in the zero-temperature limit, the Hxc potential
$v_{\rm Hxc}^{\rm SSM}(n)$ has a step of height $U$. This almost discontinuous
feature at low temperatures crosses over to a smooth one at high temperatures.
Hence  we will
use the analytic form of $v_{\rm Hxc}^{\rm SSM}$ to mimic the step feature in
our ``lowest level'' parametrization of the TBA Hxc potential. We adopt the
simplest possible way to reproduce the correct low temperature amplitude of
the step. Namely, in the function $v_{\rm Hxc}^{\rm SSM}(n,U,T)$, defined in
Eqs.~(\ref{v_hxc_ssm})-(\ref{gfunc_ssm}), the parameter $U$ will be replaced
by the zero-temperature Mott-Hubbard gap $\Delta_{0}(U)$ of
Eq.~(\ref{hubbard_gap}). The reduction of the gap automatically reduces the
value of the potential at $n=2$ from the exact value of $U$ down to
$\Delta_{0}(U)$. This unwanted behavior is corrected by adding a proper
linear function that also ensures the right point symmetry of the Hxc
potential. Putting all these arguments together we propose the following
simple  zero-level parametrization for the TBA Hxc potential (p0TBA)
\be
v_{\rm Hxc}^{\rm p0TBA}(n) = \frac{U-\Delta_{0}(U)}{2} n +
v_{\rm Hxc}^{\rm SSM}(n,\Delta_{0}(U),T).
\label{vhxc_tba_para0}
\ee
The Hxc potential defined by this equation is shown in
Fig.~\ref{vhxc_tba_comp} together with the full numerical
$v_{\rm Hxc}^{\rm TBA}(n)$. We clearly see that for all $T$ and $U$ the
parametrization proposed in Eq.~(\ref{vhxc_tba_para0}) overall
agrees reasonably well with $v_{\rm Hxc}^{\rm TBA}(n)$. The maximal deviations
never exceed unity (i.e., the value of the hopping integral $t$). Obviously
the simple
linear form of the first term of Eq.~(\ref{vhxc_tba_para0}) is
not flexible enough to reproduce the detailed features of the full TBA Hxc
potential away from the step (see Fig.~\ref{vhxc_tba_comp}). However, as long
as the difference between the parametrization and the full TBA Hxc potential
are small compared to $t$, these inaccuracies are, in most practical cases, of
little consequence for the solution of the KS equations, as will be confirmed
in Sec.~\ref{applic}. It is also worth noting that a reasonably
accurate practical approximation of Eq.~(\ref{vhxc_tba_para0}) does not
actually require the solution of the TBA equation. The only input we used was a
zero-temperature Mott-Hubbard gap and general symmetry arguments. This
observation can be useful to construct local approximations for more
complicated, e.~g. multidimensional, lattice models for which no exact
solutions are available.

Apparently there are cases when the fine structure of the density distribution
cannot fully be captured within our simplest zero-level parametrization
p0TBA defined by Eq.~(\ref{vhxc_tba_para0}). Therefore it is desirable to
design a refined parametrization which further reduces the deviation from the
numerical TBA potential. Two successive refinements of the ``first-level''
(p1TBA), and of the ``second level'' (p2TBA) are described in the next two
subsections.

\subsection{First-level refined parametrization correcting the temperature
dependence: p1TBA}

Fig.~\ref{vhxc_tba_comp} clearly shows that the temperature dependence of our
simple p0TBA potential Eq.~(\ref{vhxc_tba_para0}) is not perfect. At higher
temperatures the step in the function $v_{\rm Hxc}^{\rm p0TBA}(n)$ washes out
too fast as compared to the numerical $v_{\rm Hxc}^{\rm TBA}(n)$. There is an
obvious physical reason for this deficiency. When the temperature $T$
increases and becomes larger than unity (in units of the hopping integral
$t$), the kinetic energy contribution to the partition function becomes less
and less important. Therefore at $T>1$, and independently of $U$,
the system should behave more or
less like a collection of independent sites with the Hxc potential given by
the pure SSM expression of Eqs.~(\ref{v_hxc_ssm})-(\ref{gfunc_ssm}).

In our first-level refinement (p1TBA) we take into account this physics by
replacing $\Delta_0(U)$ in Eq.~(\ref{vhxc_tba_para0}) with a
``temperature-dependent gap'' $\Delta_1(U,T)$
\be
v_{\rm Hxc}^{\rm p1TBA}(n) = \frac{U-\Delta_1(U,T)}{2} n +
v_{\rm Hxc}^{\rm SSM}(n,\Delta_1(U,T),T) .
\label{vhxc_tba_para1}
\ee
The function $\Delta_1(U,T)$ reduces to $\Delta_0(U)$ at $T\ll 1$ and
approaches $U$ in the opposite limit of $T\gg 1$. The two limits are
connected by a smooth function which is determined by comparison with
the numerical TBA data. We have found that the following Pad\'{e}-like form does
the required job
\begin{eqnarray}
\Delta_1(U,T)=\frac{\Delta_0(U)+a^{(1)}(T) T+U T^2}{1+T^2}
\label{Delta1_T}
\end{eqnarray}
with
\[
a^{(1)}(T)=\frac{a_1^{(1)} T+a_2^{(1)}}{T^2+a_3^{(1)}}\,.
\]
Here $a_1^{(1)}=0.95, a_2^{(1)}=-0.08, a_3^{(1)}=0.13$.

\begin{figure}[t]
\includegraphics[width=0.47\textwidth]{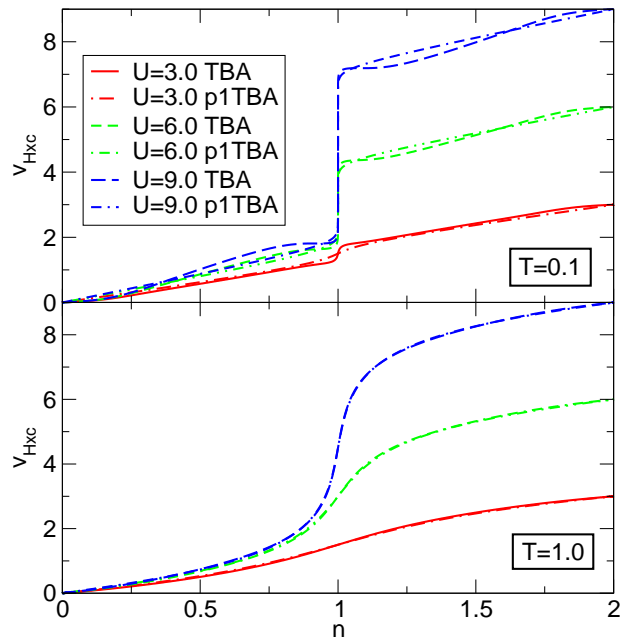}
\caption{Comparison of the fully numerical TBA Hxc potential with the
first-level refined p1TBA parametrization of
Eq.~(\protect\ref{vhxc_tba_para1}) for two typical temperatures.}
\label{vhxc_tba_comp1}
\end{figure}

From Fig.~\ref{vhxc_tba_comp1} we see that the first-level parametrization
p1TBA, Eqs.~(\ref{vhxc_tba_para1})-(\ref{Delta1_T}), produces an Hxc potential
which is practically indistinguishable from the full numerical
$v_{\rm Hxc}^{\rm TBA}(n)$, provided that $T$ is not too small. However, there
are still some deviations in the low-temperature regime. This point is
addressed at the last step in our three-level hierarchy of parametrizations.

\subsection{Second-level refinement -- the best analytic fit to numerical TBA:
p2TBA}
\label{IVC}

\begin{figure}[t]
\includegraphics[width=0.47\textwidth]{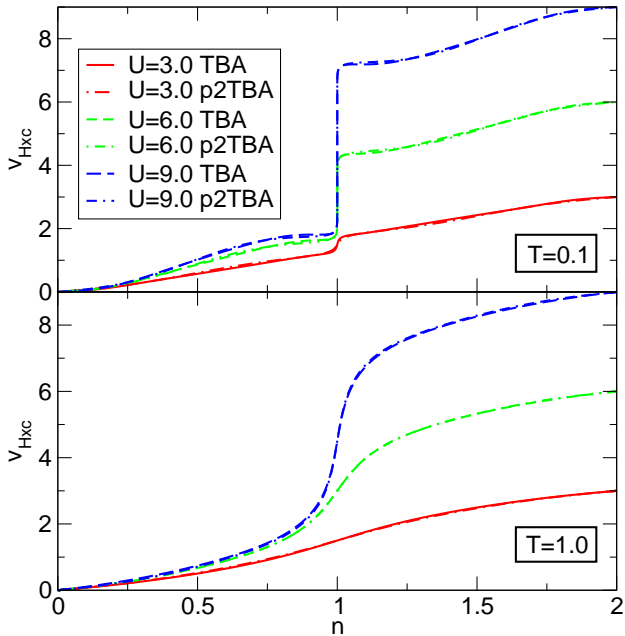}
\caption{Comparison of the fully numerical TBA Hxc potential with the
second-level refined p2TBA parametrization of
Eq.~(\protect\ref{vhxc_tba_para2}) for two typical temperatures.}
\label{vhxc_tba_comp2}
\end{figure}

The only feature missing in the first-level p1TBA parametrization is a
low-temperature nonlinearity of $v_{\rm Hxc}^{\rm TBA}(n)$ away from the half
filling (see upper panel of Fig.~\ref{vhxc_tba_comp1}). Physically the
nonlinearity should be attributed
to a nontrivial density of states in the Hubbard bands. According to our
experience the remaining inaccuracy in the Hxc potential has practically no
effect on the density distribution. On the other hand, we cannot exclude
that in some situations (steep external potentials and/or small number of
particles) the low-temperature inaccuracy of $v_{\rm Hxc}^{\rm p1TBA}(n)$ will
produce visible (thought definitely not large) errors in the density. To
avoid such problems we go to the last step in our hierarchy and introduce a
nonlinear correction term. A very satisfactory fit to the numerical
$v_{\rm Hxc}^{\rm TBA}(n)$ can be achieved with the following (p2TBA) form
\begin{eqnarray}
\lefteqn{
v_{\rm Hxc}^{\rm p2TBA}(n) = \frac{U-\Delta_2(U,T)}{2} n +
v_{\rm Hxc}^{\rm SSM}(n,\Delta_2(U,T),T)} \nn\\
&&- A(U,T)\sin[2\pi(n-1)] -  B(U,T)\sin[\pi(n-1)] .
\label{vhxc_tba_para2}
\end{eqnarray}
We note that the analytic form of the
correction term in Eq.~(\ref{vhxc_tba_para2}) automatically preserves the
point symmetry of the potential and the exact values at the end points,
$n=0$ and $n=2$. Note also that we use a new function $\Delta_2(U,T)$ defined
by
\begin{eqnarray}
\Delta_2(U,T)=\frac{\Delta_0(U)+a^{(2)}(T) T+U T^2}{1+T^2}
\label{Delta2_T}
\end{eqnarray}
with
\[
a^{(2)}(T)=\frac{a_1^{(2)} T+a_2^{(2)}}{T+a_3^{(2)}}
\]
and the coefficients are $a_1^{(2)}=-0.28$, $a_2^{(2)}=2.2$, and
$a_3^{(2)}=0.50$. As before, in the zero-temperature limit $\Delta_2(U,T)$
reduces to the correct zero-temperature gap $\Delta_0(U)$ while in the
high-temperature limit it becomes $U$. The interaction and temperature
dependent coefficients $A(U,T)$ and $B(U,T)$ in Eq.~(\ref{vhxc_tba_para2}) are
parametrized as follows
\begin{eqnarray}
A(U,T)=\frac{A_1(T) U^2}{U^2+A_2(T)},\,B(U,T)=\frac{B_1(T)U^2}{U^2+B_2(T)}\,,
\label{a_b_coeffs}
\end{eqnarray}
with
\begin{eqnarray}
A_1(T)=\frac{A_{11}}{T^2+A_{12}},\,A_2(T)=\frac{A_{21}}{T^2+A_{22}}\,,
\end{eqnarray}
\begin{eqnarray}
B_1(T)=\frac{B_{11}}{T^2+B_{12}},\,B_2(T)=\frac{B_{21}}{T^2+B_{22}}\,.
\end{eqnarray}
Here $A_{11}=0.09, A_{12}=0.25, A_{21}=705.5, A_{22}=24.95$ and $B_{11}=0.05,
B_{12}=0.13, B_{21}=0.65, B_{22}=0.01$. The accuracy of this parametrization
can be appreciated in Fig.~\ref{vhxc_tba_comp2}: at the scale of the plot,
the parametrization p2TBA and the full numerical TBA Hxc potentials are
essentially indistinguishable.

Our parametrization of the finite-temperature TBA results
generalizes earlier parametrizations
\cite{LimaSilvaOliveiraCapelle:03,FrancaVieiraCapelle:12} valid for zero
temperature. Close comparison of the Hxc potential of our p2TBA
parametrization of Eq.~(\ref{vhxc_tba_para2}) in the zero-temperature limit
with the FVC parametrization of Ref.~\onlinecite{FrancaVieiraCapelle:12}
and exact zero-temperature results reveals that the p2TBA parametrization
in some density ranges can be marginally less accurate than FVC. As pointed
out before, however, in contrast to FVC our parametrization by construction
incorporates the exact zero-temperature gap. Since the most prominent feature
of the Hxc potential as function of density is precisely the discontinuity
at half-filling, i.e., the zero-temperature gap, in some situations the
incorrect gap of FVC might lead to spurious features as will be shown below.

\section{Self-consistency with rapidly varying functionals using bisection}
\label{conv}

In the previous Section we presented a hierarchy of explicit local
approximations for the Hxc potential of lattice DFT, which
at low temperatures are
rapidly varying functions of the density close to half-filling. In
Sec.~\ref{KS_problem} we have pointed out the difficulties in converging
the usual self-consistency cycle for solving the KS equation
(\ref{ks_eq_general}) with such functionals. In the present Section we show
how to avoid the convergence problem and present a numerically feasible
algorithm to obtain the self-consistent solution based on bisection techniques.

We begin by writing again the self-consistency equation for the density at
site $i$, Eq.~(\ref{dens}), in a form of a fixed point problem, i.~e., making
explicit its dependence on the densities at all other sites:
\be
n_i = \sum_{\alpha} f(\varepsilon_{\alpha}(\vn))
| \varphi^{(\alpha)}_i(\vn)|^2 \equiv G_i[\vn]
\label{dens_dens}
\ee
where
\be
\vn = (n_1,n_2, \ldots,n_{N_s})
\ee
and the orbitals are calculated from the KS equation (\ref{ks_eq}) using
the KS potential
\be
v_i^{\rm KS}(n_i) = v_i + v_{\rm Hxc}^{\rm mod}(n_i) \; .
\label{dens_KS}
\ee
The set of equations (\ref{dens_dens}) for $i\in {1, \ldots,N_s}$ constitute a
coupled set of $N_s$ nonlinear equations for the $N_s$ ground-state densities
$n_i$, $i=1,\ldots,N_s$. As we have argued in Sec.~\ref{KS_problem}, the
numerical solution of these equations by plain iterations produces a
non-converging sequence. Obviously more elaborate iterative schemes,
such as the Newton-Rhapson method, which incorporates information on derivatives
of the equations with respect to the unknown variables are also not appropriate
because those derivatives may become very large (in the low-temperature
regime). This again leads to convergence problems in the iterative
solution of the coupled nonlinear equations.

Here we propose a solution scheme based on bisection. The main idea of our
algorithm is inspired by the exactly solvable single-site KS problem described
in Sec.~\ref{SSM}. Therefore we first explain it for this simple, but very
illuminating case. Instead of iteratively searching for the fixed point of
function $G(n)$ in Eq.~(\ref{fixed_point_ssm}) we rewrite this equation as
\begin{equation}
 \label{bisec_ssm}
n - G(n) = 0,
\end{equation}
and search for zeros of the left hand side. From the uniqueness of the
solution we know that there is only one zero, and, by construction, the l.h.s.
of Eq.~(\ref{bisec_ssm}) has different signs at the end points, $n=0$ and
$n=2$, of the density interval. Therefore the standard bisection method
\cite{PressTeukolskyVetterlingFlannery:86} is applicable. Starting from the
end points and using bisections we can bracket the solution to any desired
accuracy.

For the general lattice KS problem we need to solve a system of
Eqs.~(\ref{dens_dens}). In this case the following straighforward
multidimensional generalization of the standard bisection method can be
used. We start with an initial guess $n_i^{(0)}$ for the densities at
sites $i \in\{2, \ldots, N_s\}$. We then define the density vector
\be
\vn^{(1)} = \left( n_1, n_2^{(0)}, \ldots, n_{N_s}^{(0)}\right) \;,
\label{dens_init}
\ee
insert this density vector in the rhs of Eq.~(\ref{dens_dens}) for $i=1$ and
solve the resulting nonlinear equation for the density $n_1=n_1^{(1)}$ with the
usual 1D bisection method. Then we choose the density vector
\be
\vn^{(2)} = \left( n_1^{(1)},n_2, n_3^{(0)}, \ldots, n_{N_s}^{(0)}\right) \;,
\ee
insert it into Eq.~(\ref{dens_dens}) for $i=2$ and solve for $n_2=n_2^{(1)}$
again by bisection. In the next step we take
\be
\vn^{(3)} = \left( n_1^{(1)},n_2^{(1)}, n_3, n_4^{(0)}, \ldots, n_{N_s}^{(0)}
\right)
\ee
and solve Eq.~(\ref{dens_dens}) for $i=3$ for $n_3=n_3^{(0)}$. We continue
the procedure until we have exhausted the $N_s$ equations (\ref{dens_dens}).
Then we start the cycle all over but now with the initial guess for the
density $\vn^{(N_s+1)}$ as Eq.~(\ref{dens_init}) but with the $n_i^{(0)}$
replaced by $n_i^{(1)}$. The whole process is continued until convergence is
achieved.

\begin{figure}[t]
\includegraphics[width=0.5\textwidth]{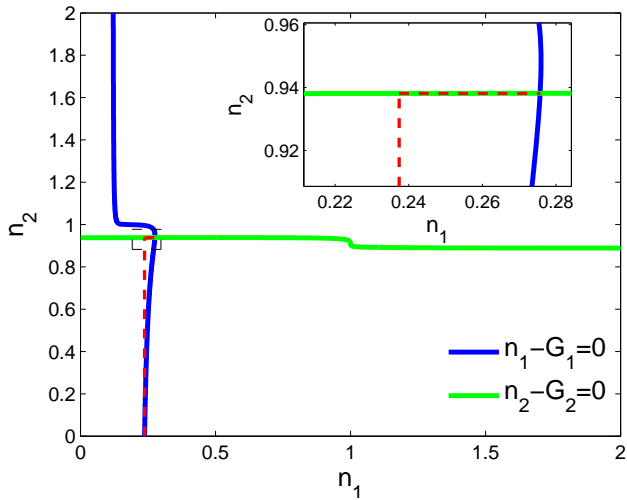}
\caption{Solution lines of $n_i-G_i(n_1,n_2)=0$, $i=1,2$, for the symmetric
three-site problem. The intersection of these two lines is the solution to
the system of Eqs.~(\ref{bisec_site1})-(\ref{bisec_site2}).  The parameters
are $U=8$, $\mu=0.5$, $v_1=2$, $v_2=0$, and $T=0.5$.}
\label{n1n2}
\end{figure}

In Fig.~\ref{n1n2} we illustrate the iterative procedure for a symmetric
three-site problem, i.e., for $v_1=v_3$ and by symmetry also
$v_1^{\rm KS}=v_3^{\rm KS}$. We then need to solve two coupled nonlinear
equations of the form
\begin{eqnarray}
 \label{bisec_site1}
n_1 &-& G_1(n_1,n_2)= 0, \\
 \label{bisec_site2}
n_2 &-& G_2(n_1,n_2)= 0.
\end{eqnarray}
In Fig.~\ref{n1n2} we show the $(n_1,n_2)$ plane and zero lines for the l.h.s.
of Eqs.~(\ref{bisec_site1}) and (\ref{bisec_site2}). The overall solution of
the problem is given by the intersection of these two lines. Furthermore, the
dashed line in Fig.~\ref{n1n2} shows how our iterative scheme converges to
this solution.

From this example we can also understand that in some cases the proposed
scheme, as is, may not converge. If the zero lines close to the intersection
become just straight lines then, depending on the slopes of these lines, the
iterative scheme may follow a rectangular path encircling
the solution point but never reaching it. However, in this case one additional
Newton-Rhapson step (using information on the derivatives) will directly lead
to the solution point.

As is common in DFT, we have used the densities as fundamental variables
in Eq.~(\ref{dens_dens}). It is also possible to implement the bisection
scheme in terms of the KS potentials. To this end we write the vector of
Hxc potentials at sites $i$ as
\be
\vv_{\rm Hxc} = \left( v_{{\rm Hxc},1}, \ldots, v_{{\rm Hxc},N_s}\right)
\label{vfunctional1}
\ee
and consider both the KS orbitals and eigenvalues as functions of this vector,
i.e., $\varphi^{(\alpha)}=\varphi^{(\alpha)}(\vv_{\rm Hxc})$ and
$\varepsilon^{(\alpha)}=\varepsilon^{(\alpha)}(\vv_{\rm Hxc})$. Therefore, also
the density at site $i$ can be considered a function of $\vv_{\rm Hxc}$ through
\be
n_i(\vv_{\rm Hxc}) = \sum_{\alpha} f(\varepsilon_{\alpha}(\vv_{\rm Hxc}))
| \varphi^{(\alpha)}_i(\vv_{\rm Hxc})|^2 \;.
\label{dens_vhxc}
\ee
For our local approximations to the Hxc potentials the set of nonlinear
equations to be solved by bisection then becomes
\be
v_{{\rm Hxc},i} = v_{\rm Hxc}^{\rm mod}(n_i(\vv_{\rm Hxc}))
\label{vfunctional3}
\ee
with the local density $n_i(\vv_{\rm Hxc})$ given by Eq.~(\ref{dens_vhxc}).
In order to solve this set of equations we proceed in an analogous way to
the one described above for the density-based procedure but now treating the
local Hxc potentials $v_{{\rm Hxc},i} $ at site $i$ as the unknowns to be
determined.

\section{Numerical applications}
\label{applic}

In this Section we present some numerical examples demonstrating successful
applications of our self-consistency algorithm as well as the accuracy of the
local approximation using our different parametrizations.

\begin{figure}[t]
\includegraphics[width=0.47\textwidth]{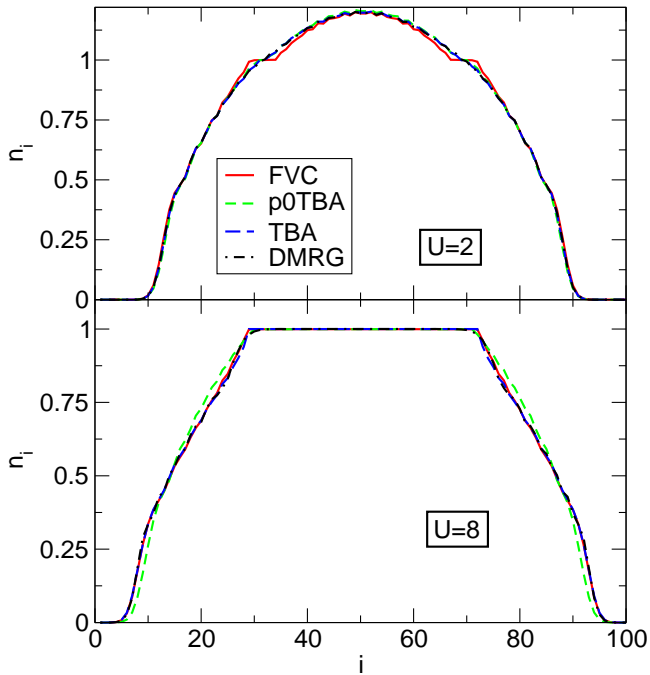}
\caption{Density distribution of $N=70$ particles in the harmonic potential at
$T=0$ (upper panel: $U=2$, lower panel: $U=8$). The strength of the external
potential is $V_{\rm ext}=2.5 \times 10^{-3}$. We compare the exact DMRG
results against DFT results based on different choices for $v_{\rm Hxc}$.
They are the parametrization of Ref.~\onlinecite{FrancaVieiraCapelle:12}
(labeled as FVC), our lowest-level parametrization p0TBA of
Eq.~(\ref{vhxc_tba_para0}) (labeled as p0TBA), and the full numerical TBA
potential (labeled as TBA), respectively. The results obtained with the p2TBA
potential of Eq.~(\ref{vhxc_tba_para2}) are indistinguishable from the full
TBA results.}
\label{dens_harmonic_trap}
\end{figure}

We illustrate our theoretical developments by calculating the density
distribution of particles confined by an external harmonic potential of the
form
\be
\label{trap}
v_i = V_{\rm ext} \; (i-i_0)^2
\ee
where $V_{\rm ext}$ is the strength of the trapping
potential and $i_0=(N_s+1)/2$ (we take $V_{\rm ext}$ in units of the hopping 
parameter $t$). The
Hubbard chain with a superimposed harmonic potential is commonly used to
model the behavior of cold fermionic gases in 1D optical lattices
~\cite{RigolMuramatsuBatrouniScalettar:03,LiuDrummondHu:05,CampoCapelle:05,XianlongPoliniTanatarTosi:06,Heiselberg:06,YamamotoYamadaOkumuraMachida:11}.
Therefore our results
below have a clear relevance for the physics of cold trapped atoms. However,
for our present illustrative purposes, the choice of this particular system
is related to one of its specific features, namely the possible coexistence
of the Mott insulator phase around the center of the trap and the metallic
phase at the trap's edges~\cite{RigolMuramatsuBatrouniScalettar:03,LiuDrummondHu:05,XianlongPoliniTosiCampoCapelleRigol:06}. In the Mott phase the
density is pinned at $n=1$ which shows up as an extended plateau in the
density distribution. Therefore in trapped systems the appearance of the
Mott insulator phase becomes detectable within the ``density-only`` DFT
concept. On the other hand, at the level of DFT functionals the Mott physics
is solely related to the discontinuity of the xc potential. Thus the Hubbard
model with a harmonic confinement is perfectly suited for demonstrating the
working power of our algorithm as well as the performance of different
parametrizations for the Hxc potentials.

As a first example we study a system with $N=70$ particles on
$N_s=100$ sites in the presence of the potential given by Eq.~(\ref{trap}) 
with $V_{\rm ext}=2.5\times 10^{-3}$.
In Fig.~\ref{dens_harmonic_trap} we show self-consistent
densities for two different values of the Hubbard interaction, $U=2$ and
$U=8$, evaluated at zero temperature. Except for the DMRG results which denotes
numerically exact reference results from density matrix renormalization group 
calculations~\cite{White:92,Schollwoeck:05,Albuquerque:07}, all
other calculations result from self-consistent DFT calculations with local
approximations to the Hxc potential. FVC denotes the zero-temperature BALDA
using the parametrization of Ref.~\onlinecite{FrancaVieiraCapelle:12}, p0TBA
denotes results obtained with our low-level parametrization of
Eq.~(\ref{vhxc_tba_para0}), the results obtained from our second-level
parametrization of Eq.~(\ref{vhxc_tba_para2}) are indistinguishable from those
using the exact numerical TBA as input in the local approximation.

\begin{figure}[t]
\includegraphics[width=0.45\textwidth]{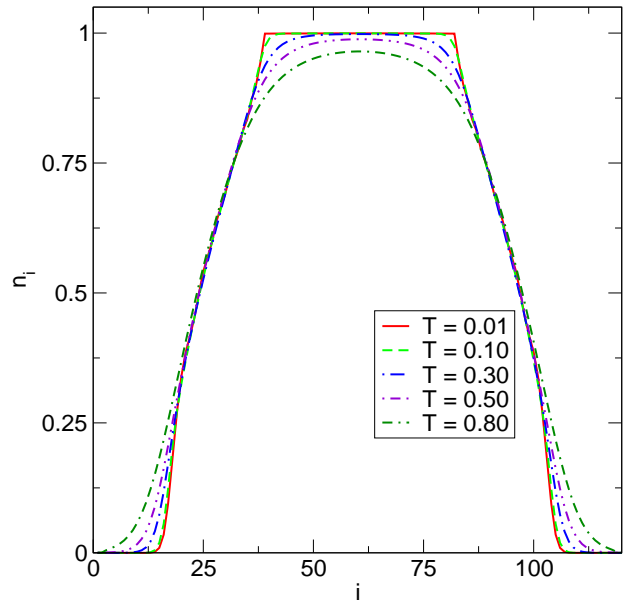}
\caption{Melting of the Mott phase at finite temperature. The calculations
are based on the p2TBA parametrization which produces the results identical
to those of the full numerical TBA.
The parameters are $U=8$, $V_{\rm ext}=2.5 \times 10^{-3}$, $N=70$.}
\label{T_effect}
\end{figure}

We see that the density profiles are quite similar in the different
approaches. For $U=2$ the FVC parametrization exhibits two spurious density
plateaus around $i\approx30$ and $i\approx70$ which are due to the fact that
this parametrization does not incorporate the zero-temperature derivative
discontinuity exactly but only approximately. We also see some small
differences between the p0TBA and the TBA results in the flanks of the density
profile. For stronger interaction, $U=8$, the density exhibits an extended
plateau of value unity over a wide range of sites in the center of the well
which physically corresponds to the local, incompressible Mott phase.
On the other hand, in the density functional picture this plateau is a direct
consequence of the extremely rapid variation of the Hxc potentials as function
of the density. The small difference in the p0TBA and TBA densities is
related to the nonlinearity of Hxc potential away from the half filling
(see Sec.~\ref{IVC}). This deficiency is corrected in our second level
parametrization p2TBA of Eq.~(\ref{vhxc_tba_para2}). As a result the density
calculated with the p2TBA Hxc potential is completely indistinguishable from
that obtained using the full numerical TBA potential. This also holds true for
all interactions, temperatures and trapping potentials we have tried. Hence in
practice in all figures the results denoted as TBA have been actually produced
using the p2TBA potential defined after Eq.~(\ref{vhxc_tba_para2}).
At strictly zero temperature, the KS potential has a real discontinuity at 
integer filling $n=1$ and, therefore, it is undefined in this point. For 
practically solving Eqs. (\ref{dens_dens})-(\ref{dens_KS}) in this case we 
replace the discontinuity by a linear function when 
$n\in [1-\Delta n, 1+\Delta n]$ with $\Delta n=10^{-3}$.
Further decreasing $\Delta n$ will not hamper the convergency and the final 
results.
In general, the densities from the DFT calculations are remarkably
close to the numerically exact quantum Monte-carlo/DMRG result
\cite{RigolMuramatsuBatrouniScalettar:03,XianlongPoliniTosiCampoCapelleRigol:06,HuWangXianlongOkumuraIgarashiYamadaMachida:10}.

Unlike earlier work \cite{LimaSilvaOliveiraCapelle:03,FrancaVieiraCapelle:12},
our parametrization is valid for arbitrary temperature and thus allows to
study finite temperature effects. As a first application of this feature, in
Fig.~\ref{T_effect} we show how the density plateau due to the local Mott
phase ``melts away'' when increasing the temperature.
\begin{figure}[t]
\includegraphics[width=0.47\textwidth]{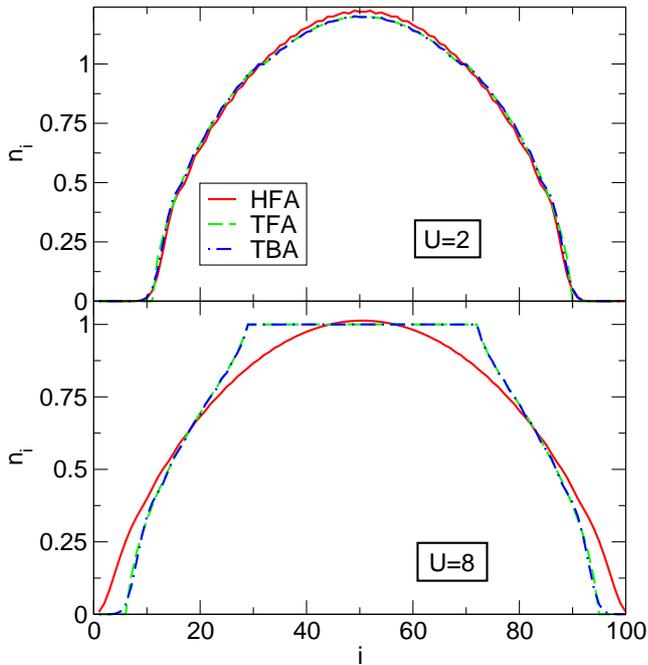}
\caption{Comparison of the density distributions at $T=0$ obtained using
Hartree-Fock approximation (HFA), Thomas-Fermi approximation (TFA) against
DFT results based on the TBA potential (TBA). The parameters are the same as
in Fig.~\ref{dens_harmonic_trap}.}
\label{dens_harmonic_trap_HFA}
\end{figure}

We have also calculated (Fig.~\ref{dens_harmonic_trap_HFA}) the density
profiles at zero temperature in the Hartree-Fock (HFA) and Thomas-Fermi
approximation (TFA). Here, by TFA we mean that the non-interacting kinetic
energy is not treated exactly as in the Kohn-Sham scheme but at the level of a
local approximation. It is important to note, however, that
exchange-correlation effects are also included at the level of the local
density approximation (in
Ref.~\onlinecite{XianlongPoliniTosiCampoCapelleRigol:06} this approximation
has been denoted as ``total-energy LDA (TLDA)''). The top panel of
Fig.~\ref{dens_harmonic_trap_HFA} shows that for small values of $U$ both the
HFA and the TFA give a reasonably accurate density profile when compared to
DMRG. For larger values of $U$ ($U=8$, lower panel of
Fig.~\ref{dens_harmonic_trap_HFA}), however, the situation is different: HFA
completely misses the development of the density plateau while TFA (due to the
local Hxc potential) does exhibit this plateau. Moreover, in HFA the density
is more spread out as compared to the exact result.

\begin{figure}[t]
\includegraphics[width=0.47\textwidth]{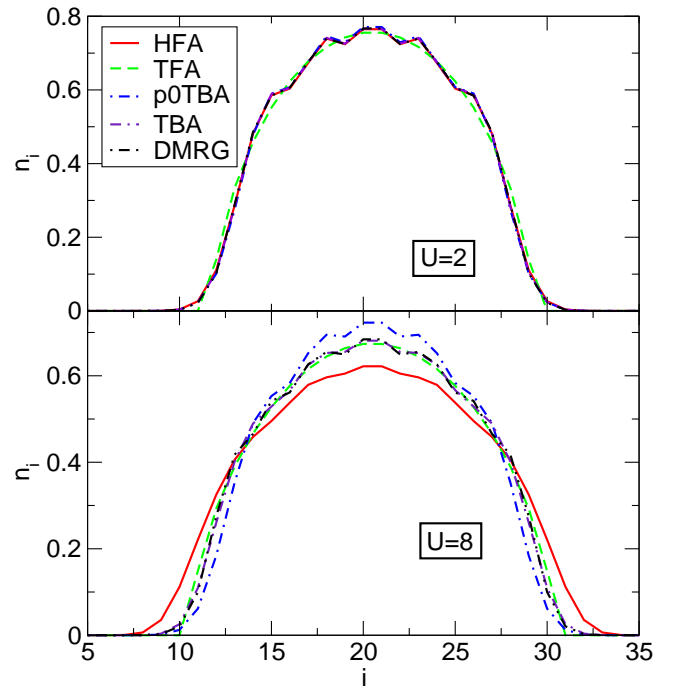}
\caption{Density distribution of $N=10$ particles in a harmonic trap at
temperature $T=0$ in Hartree-Fock (HFA) and Thomas-Fermi approximations (TFA),
compared to DFT results based on TBA and DMRG. Upper panel: $U=2$, lower
panel:  $U=8$. The external potential is $V_{\rm ext}=2.5 \times 10^{-2}$.}
\label{dens_harmonic_trap_N10}
\end{figure}

\begin{figure}[t]
\includegraphics[width=0.45\textwidth]{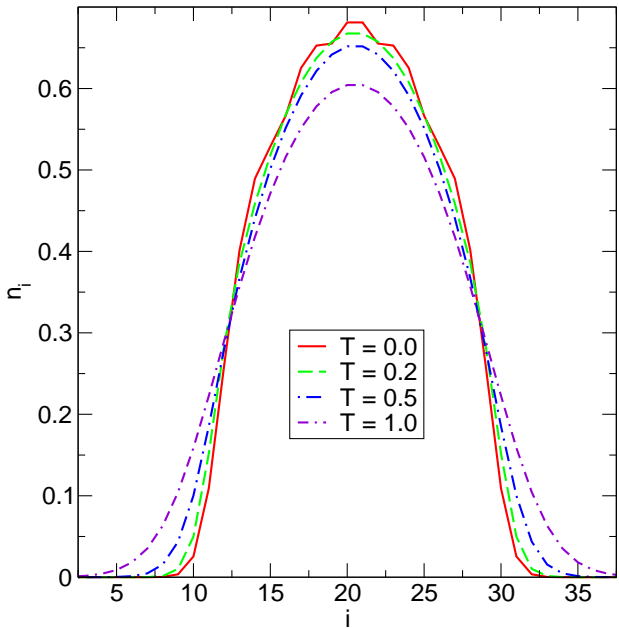}
\caption{Effects of finite temperatures on the density profile of $N=10$
particles in a harmonic trap. The calculation is based on the full numerical
TBA. The parameters are $U=8$, $V_{\rm ext}=2.5 \times 10^{-2}$.}
\label{T_effect_N10}
\end{figure}

The differences between TFA and both the DMRG as well as the full KS results
with the TBA functional are more pronounced for smaller number of particles.
In Fig.~\ref{dens_harmonic_trap_N10} we show the density profile for $N=10$
electrons on $N_s=40$ sites in the harmonic external potential of 
Eq.~(\ref{trap}) with $V_{\rm ext}=2.5\times 10^{-2}$. Not surprisingly, the 
TFA approximation completely misses the quantum oscillations in the density
profile both for small and large values of $U$. In contrast, HFA captures
these oscillations well for small $U$ since the kinetic energy is treated
at a quantum level. In contrast, for $U=8$ the correlation effects are too
strong to be captured by HFA which, again, shows a density distribution which
is too spread out. On the other hand, the KS calculation using the TBA again
is in extremely good agreement with the DMRG results. The lower panel of
Fig.~\ref{dens_harmonic_trap_N10} also shows that for this case our simplest
p0TBA parametrization of the TBA results can sometimes lead to inaccuracies.
Finally, in Fig.~\ref{T_effect_N10}, solving the KS equations with the TBA
functional, we show the effect of temperature on the density profile: one
can see that the quantum oscillations due to the relatively small number of
particles seen for zero temperature are quickly suppressed when increasing the
temperature.

\section{Conclusions}
\label{conclus}

In this work we have suggested a way to deal with convergence problems in
self-consistent KS calculations when dealing with (local) approximations for
the xc potential which exhibit rapid variations as function of the density.
Working in the framework of lattice DFT, we have formulated the KS
self-consistency
cycle as a fixed point problem and shown that for rapidly varying functionals
the fixed point in the usual prcedure is not attractive. Instead, we rephrased
the search for the self-consistent KS potential in terms of finding the roots
of a set of coupled nonlinear equations. To find these solutions, we then
suggested an iterative algorithm based on successive application of the
well-known bisection method for finding roots of nonlinear equations
in one dimension. The scheme has been successfully tested for model systems
of electrons in a harmonic trap interacting via a Hubbard interaction. We
have used a newly designed local approximation for the xc functional based on
the thermodynamic Bethe ansatz solution of the uniform Hubbard model. Based on
these results  we constructed simple, yet accurate parametrizations for
arbitrary temperatures, thus generalizing earlier parametrizations suggested
for zero temperature. This paves the way for further investigations on the
performance of finite-temperature DFT for one-dimensional lattice models.

\begin{acknowledgments}
Gao X. and A.-H. Chen were supported by the NSF of China under
Grants No. 11174253 and No. 10974181 and by the Zhejiang
Provincial Natural Science Foundation under Grant No.
R6110175.
I.V.T. and S.K. acknowledge funding by the ``Grupos Consolidados UPV/EHU del
Gobierno Vasco'' (IT-319-07) and Spanish MICINN (FIS2010-21282-C02-01).
\end{acknowledgments}

\appendix*
\section{Hartree-exchange-correlation free energy per site for the
TBA parametrizations}
\label{append}

In Sec.~\ref{lda_fintemp} we have suggested a local density approximation
for one-dimensional lattice systems based on the numerical solution of the
TBA for the uniform Hubbard model. The parametrizations we proposed were
constructed from insights gained on the numerical results for the Hxc
potential. However, typically the construction of local DFT approximations
starts by modelling the xc energy per site and the corresponding
xc potential is then obtained by differentiation. In this Appendix we derive
the expressions for the Hxc free energies per site for the
different parametrizations suggested in Sec.~\ref{lda_fintemp}.

We start with the derivation of the exact LDA Hxc free energy per site at
finite temperature, expressed in terms of general thermodynamic quantities.
As usual, a crucial ingredient of the general construction of LDA is a
reference system of interacting particles with uniform density for which the
grand canonical potential per site, $\Omega(\mu)$, is written as function of
the chemical potential $\mu$. The partition function for the reference system
is
\be
Z(\mu) = \exp(-\beta \Omega(\mu))
\ee
from which we can derive the density as function of $\mu$,
\be
n(\mu) = - \frac{\partial \Omega(\mu)}{\partial \mu} = \frac{1}{\beta}
\frac{\partial \ln Z(\mu)}{\partial \mu} \; .
\ee
This function can be inverted to give the chemical potential as function
of density, $\mu = \mu(n)$. By Legendre transformation we can then obtain
the free energy per site of the reference system as function of density as
\be
F(n) = \Omega(\mu(n)) + \mu(n) n \; .
\ee
We can repeat the same steps for the corresponding {\em non-interacting}
reference with the same uniform density $n$ with grand canonical potential per
site $\Omega_s(\mu_s)$ and the corresponding expression for the free energy per
site
\be
F_s(n) = \Omega_s(\mu_s(n)) + \mu_s(n) n \; .
\ee
The Hxc free energy per site is then simply given by
\bea
\lefteqn{
F_{\rm Hxc}(n) = F(n) - F_s(n) }\nn\\
&=& -\frac{1}{\beta} \ln\left( \frac{Z(\mu_s(n) + v_{\rm Hxc}(n))}
{Z_s(\mu_s(n))}\right) + n v_{\rm Hxc}(n)
\label{fxc_lda}
\eea
where we have defined the Hxc potential as
\be
v_{\rm Hxc}(n) = \mu(n) - \mu_s(n)
\ee
(see also Eq.~(\ref{vhxcTBA})).

Applying Eq.~(\ref{fxc_lda}) to the single-site model of
Section \ref{SSM}, we obtain for the Hxc free energy per site of that model
\bea
\lefteqn{
F_{\rm Hxc}^{\rm SSM}(n,U,T) = n v_{\rm Hxc}^{\rm SSM}(n,U,T) }\nn\\
&& -\frac{1}{\beta} \ln\left( \frac{Z^{\rm SSM}(\mu_s^{\rm SSM}(n,T) +
v_{\rm Hxc}^{\rm SSM}(n,U,T))}
{Z_s^{\rm SSM}(\mu_s^{\rm SSM}(n,T))}\right)
\eea
where the Hxc potential of the SSM model,
$v_{\rm Hxc}^{\rm SSM}(n,U,T))$, is given by Eq.~(\ref{v_hxc_ssm}) and we have
made explicit the dependence on the temperature $T$ and the on-site
interaction $U$. It remains
to find the dependence of $Z^{\rm SSM}$ and $Z_s^{\rm SSM}$ on the density.
This can be done by performing the program described above with the
interacting and non-interacting partition functions of the
SSM model given in Eqs.~(\ref{ssm_partfunc_int}) and
(\ref{ssm_partfunc_nonint}). Actually, we have already calculated the
dependence of the chemical potential on the density (see
Eq.~(\ref{ssm_pot_nonint}))
\be
\mu_s^{\rm SSM}(n,T) = -\tilde{v}_s(n) =
\frac{1}{\beta} \ln \left( \frac{n}{2-n}\right)
\label{mu_s_ssm}
\ee
which, when inserted back into Eq.~(\ref{ssm_partfunc_nonint}), yields
the non-interacting partition function of the SSM model in terms of the
density
\be
Z_s^{\rm SSM}(n) = \left( \frac{2}{2-n} \right)^2 \;.
\label{zs_ssm}
\ee
Finally, the interacting partition function of the SSM model (see
Eq.~(\ref{ssm_partfunc_int})) may be written in terms of the density as
\bea
\lefteqn{
Z^{\rm SSM}(n,U,T) = 1 + \frac{2n}{2-n}
\exp\left(\beta v_{\rm Hxc }^{\rm SSM}(n,U,T)\right) }\nn\\
&&+ \frac{n^2}{(2-n)^2} \exp\left(\beta (2 v_{\rm Hxc }^{\rm SSM}(n,U,T) - U )
\right)
\eea
which leads to the final result for the Hxc free energy per site of the
SSM model
\bea
\lefteqn{
F_{\rm Hxc}^{\rm SSM}(n,U,T) = n v_{\rm Hxc}^{\rm SSM}(n,U,T) }\nn\\
&& -\frac{1}{\beta} \ln\Bigg[ \left(1-\frac{n}{2}\right)^2 + n
\left(1-\frac{n}{2}\right) \exp\left(\beta v_{\rm Hxc }^{\rm SSM}(n,U,T)\right)
\nn \\
&& \;\;\;\;\;\;\;\;\;\;\;\;\;\;\;
+ \frac{n^2}{4} \exp\left(\beta (2 v_{\rm Hxc }^{\rm SSM}(n,U,T) - U )\right)
\Bigg] \; .
\label{fhxc_ssm}
\eea
By construction, the derivative of this expression with respect to the density
yields the SSM Hxc potential of Eq.~(\ref{v_hxc_ssm}).

Using this result it is now easy to express the Hxc free energies per site for
our different parametrizations of the TBA results. For the lowest-level
parametrization (p0TBA) of the TBA the resulting expression reads
\be
F_{\rm Hxc}^{\rm p0TBA}(n,U,T) = \frac{U-\Delta_0(U)}{4} n^2 +
F_{\rm Hxc}^{\rm SSM}(n,\Delta_0(U),T)
\label{fxcp0}
\ee
where $\Delta_0(u)$ is the exact zero-temperature gap of the uniform Hubbard
model given by Eq.~(\ref{hubbard_gap}). For the first-level refinement
(p1TBA), it has the same form except that $\Delta_0(U)$ is replaced by
$\Delta_1(U,T)$ of Eq.~(\ref{Delta1_T}), i.e.,
\bea
F_{\rm Hxc}^{\rm p1TBA}(n,U,T) &=& \frac{U-\Delta_1(U,T)}{4} n^2 \nn\\
&& + F_{\rm Hxc}^{\rm SSM}(n,\Delta_1(U,T),T) \; .
\label{fxcp1}
\eea
Finally, for the second refined parametrization (p2TBA) we have
\bea
\lefteqn{
F_{\rm Hxc}^{\rm p2TBA}(n,U,T) = \frac{U-\Delta_2(U,T)}{4} n^2 } \nn\\
&& + F_{\rm Hxc}^{\rm SSM}(n,\Delta_2(U,T),T) +
\frac{A(U,T)}{2 \pi} \left( \cos(2\pi(n-1)) - 1 \right) \nn\\
&& + \frac{B(U,T)}{\pi} \left( \cos(\pi(n-1)) + 1 \right)
\label{fxcp2}
\eea
with $\Delta_2(U,T)$ given by Eq.~(\ref{Delta2_T}) and the functions $A(U,T)$
and $B(U,T)$ given by Eq.~(\ref{a_b_coeffs}).

It is worth mentioning that for zero temperature, since
$\Delta_2(U,0)=\Delta_1(U,0)=\Delta_0(U)$, the contributions to the Hxc free
energy per site coming from the single-site model in all our parametrizations
(\ref{fxcp0})-(\ref{fxcp2}) reduce to the same limit
\be
F_{\rm Hxc}^{\rm SSM}(n,\Delta_0(U),0) = \Delta_0(U) (n-1) \Theta(n-1)
\ee
where $\Theta(x)$ is the Heaviside step function.


\end{document}